\documentclass[graybox]{svmult}

\usepackage{graphicx}
\usepackage{type1cm}        
\usepackage{makeidx}         
\usepackage{multicol}        
\usepackage[bottom]{footmisc}

\usepackage{newtxtext}       %
\usepackage{newtxmath}       

\makeindex             

\begin{document}

\title*{A Soft Robotic Cover with Dual Thermal Display and Sensing Capabilities}
\author{Yukiko Osawa and Abderrahmane Kheddar}
\institute{Yukiko Osawa and Abderrahmane Kheddar \at CNRS-University of Montpellier, LIRMM, Interactive Digital Humans group, Montpellier, France \email{yukiko.akiyama@lirmm.fr}}

\maketitle

\abstract*{We propose a new robotic cover prototype that achieves thermal display while also being soft. We focus on the thermal cue because previous human studies have identified it as part of the touch pleasantness. The robotic cover surface can be regulated to the desired temperature by circulating water through a thermally conductive pipe embedded in the cover, of which temperature is controlled. Besides, an observer for estimating heat from human contact is implemented; it can detect human interaction while displaying the desired temperature without temperature sensing on the surface directly. We assessed the validity of the prototype in experiments of temperature control and contact detection by human hand.}

\abstract{We propose a new robotic cover prototype that achieves thermal display while also being soft. We focus on the thermal cue because previous human studies have identified it as part of the touch pleasantness. The robotic cover surface can be regulated to the desired temperature by circulating water through a thermally conductive pipe embedded in the cover, of which temperature is controlled. Besides, an observer for estimating heat from human contact is implemented; it can detect human interaction while displaying the desired temperature without temperature sensing on the surface directly. We assessed the validity of the prototype in experiments of temperature control and contact detection by human hand. }
\thispagestyle{empty}
\section{Introduction}
\label{sec:introduction}

Physical interaction between human and robot enables various robotic tasks: helping a frail person in daily life, programming a robot in manufacturing through touch guidance and demonstration, etc. In order to achieve such tasks, robots can be conceived in order to display features such as soft texture and warmth, in order to be pleasant to human touch. We claim that the interaction between humans and robots shall also be considered from an aesthetic and pleasantness (to human touch) hardware design viewpoint. In our view, combining functionalities with aesthetics and pleasantness is an important dimension in human-robot physical interaction and worth investigating in human-centric robotics. 
Moreover, robot cover's softness is at the intersection of both functionality (casting the environment irregularities, absorbing impacts, safety through inherent compliance, etc.) and pleasantness to touch when compared to current rigid design. In the other hand, warmth in physical interactions has beneficial effects~\cite{nie2012hri,park2014robotica}. 
Therefore, combining thermal exchanges and softness properties to design robotic covers potentially increases embodiment, acceptability, and human friendliness. 

The challenge is that (i) most of the soft and low-density materials are not thermally conductive and hence cannot display changes in temperature on demand; (ii) most of the thermally conductive material are also electric conductive and have high-density mass which is problematic to cover all the robot limbs (e.g. the humanoid case); (iii) if any cover material exist, one still need to provide desired temperature display using a heat source. But integrating a heat source --as in haptic displays technologies~\cite{drif2005iros}, for the entire cover surface is neither a realistic nor a plausible perspective. We solve this issue by using flowing liquid (water) with the Peltier pump that can heat or cool it. 

Our aim is to achieve robotic covers that are both soft and capable of thermal display. Another side idea is to exploit the thermal cover in a dual thermal cues display and sensing. This would allow the robot to also detect human touch through monitoring the thermal exchange during the process of robotic cover thermal display regulation. 
Thermal display techniques are found in haptics research. Namely in portable haptic displays. Indeed, there are many kinds of thermal displays~\cite{yamamoto2004icra,drif2005iros,jones2008haptics,guiatni2008springer,osawa2018tie,goetz2020haptics} based on a Peltier pump heat source. The latter can either heat or cool the surface (e.g. human fingertips when touched) based on the Peltier effect~\cite{rowe2018thermoelectrics}.  
Yet, to our best knowledge, thermal display technology has not yet been considered to be part of the robotic cover: this is totally novel. More challenging, is to embed thermal display to a soft robot skin that appear to be a useful feature in multi-contact~\cite{bouyarmane2019springer} and also to cast human touch. 

\begin{figure}[t!]
\begin{center}
\includegraphics[width=\columnwidth]{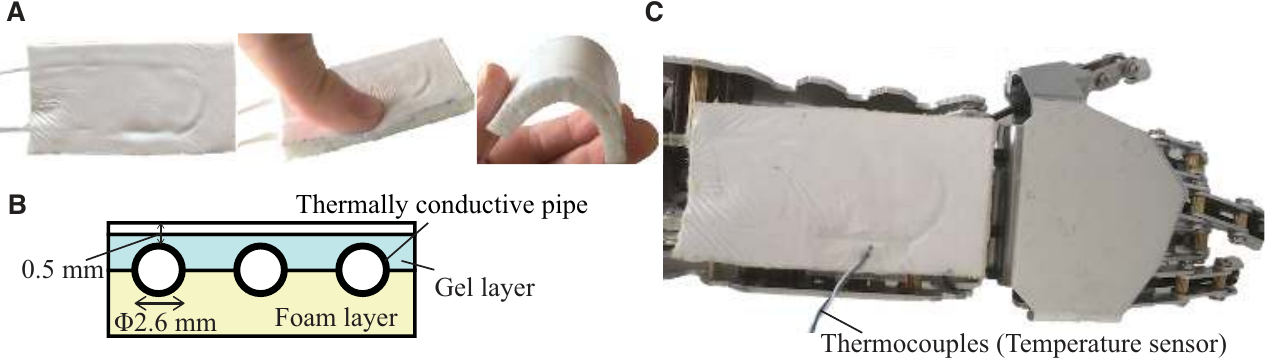}
\caption{The prototype of the robotic cover.}
\label{fig:prototype}
\end{center}
\end{figure}

\begin{figure}[t!]
\begin{center}
\includegraphics[width=0.8\columnwidth]{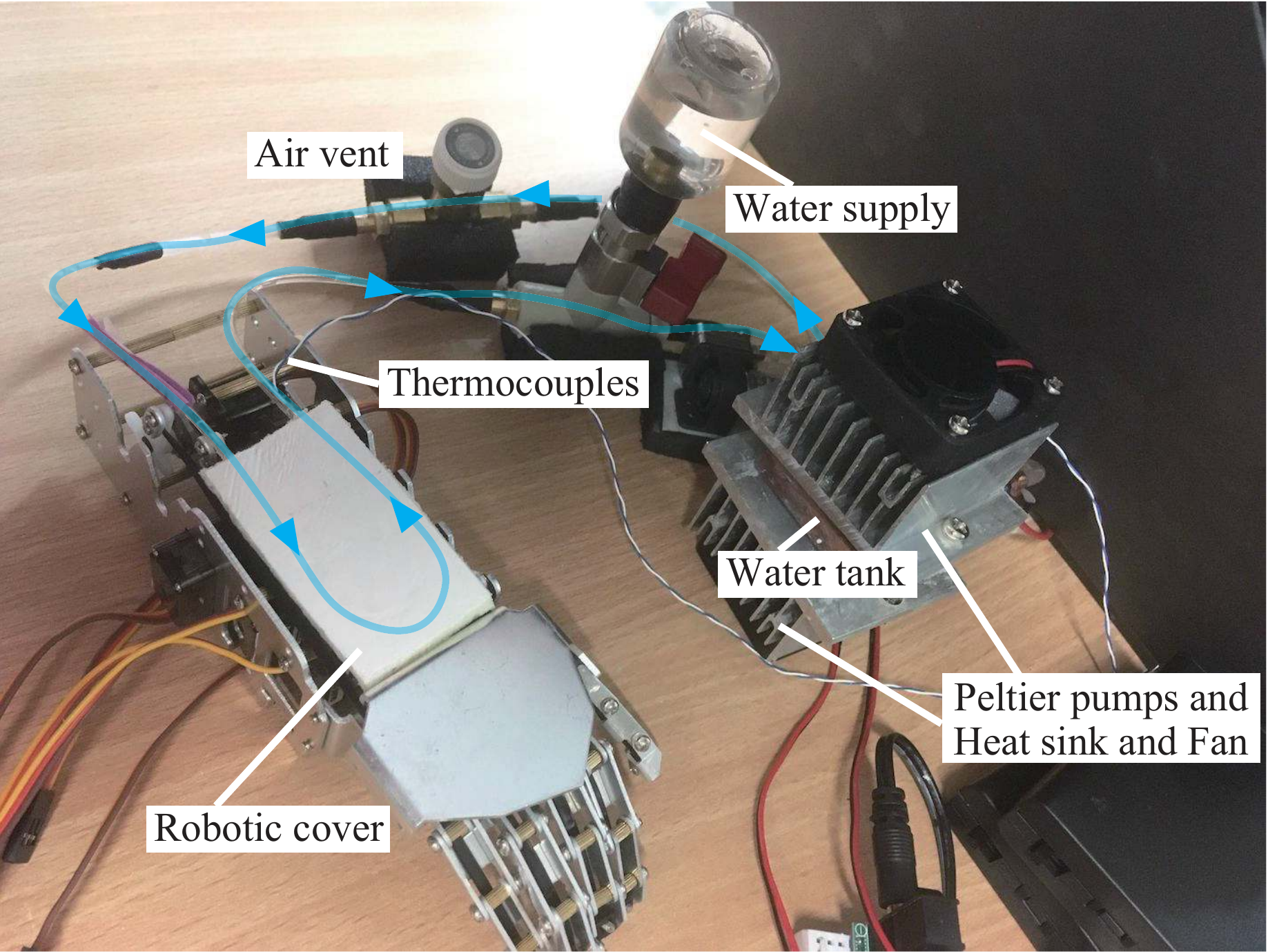}
\caption{Overall robotic cover system.}
\label{fig:robot-cover_system}
\end{center}
\end{figure}

\section{Technical Approach}
\subsection{Design of the robotic cover and its control system}
\label{sec:design}
Figure~\ref{fig:prototype}A shows the prototype of our designed composite soft and thermal robotic cover. 
The small prototype, of which size is 3.0 $\times$ 6.0~cm$^2$, is structured into two-layered materials, as shown in Fig.~\ref{fig:prototype}B. 
The upper (gel) part would conduct and hence diffuse the temperature. The foam part, being not thermally conductive, would isolate the inner robot body. 
Moreover, the chosen foam material has a very low mass density and good compliance properties, which allows designing an overall light whole-body cover that is soft with thermal rendering capabilities. These two kinds of materials consisting of the cover with high heat conductivity and low density made it possible to be soft, well thermally conductive, and light. 
Thanks to the structure, the weight of the prototype is only 4.6~g. 
For circulating water, a thermally conductive pipe is embedded between the two layers. A thin layer of gel is used to transfer temperature from the water flowing inside the pipe. The thickness of the gel and foam layer is 1~mm and 5~mm, respectively. The overall cover thickness is thus 6~mm. Ideally, we would choose the thickness to meed desired compliance of the cover while being as thin as possible. To this end, our proposed cover achieves two opposing properties: softness and high thermal conductivity at the same time. 
Some experiments were conducted by attaching the cover to the small robot hand shown in Fig.~\ref{fig:prototype}C. 

The overall system is shown in Fig.~\ref{fig:robot-cover_system}. 
It consists of a small water pump; a copper tank --on top of which is mounted a Peltier pump acting as a heat source; and finally, the soft robotic cover explained previously. The silicone pipe with a diameter of 4.5~mm, connects the water pump, the tank, and the robotic cover. The water stored in the tank is heated and cooled by the Peltier pump, regulating the circulating water temperature in a closed circuit throughout our proposed cover. Thermocouples that measure temperature are attached to the surface of the heat source, the copper water tank, and others to the soft robotic cover. 

\begin{figure}[t!]
\begin{center}
\includegraphics[width=\columnwidth]{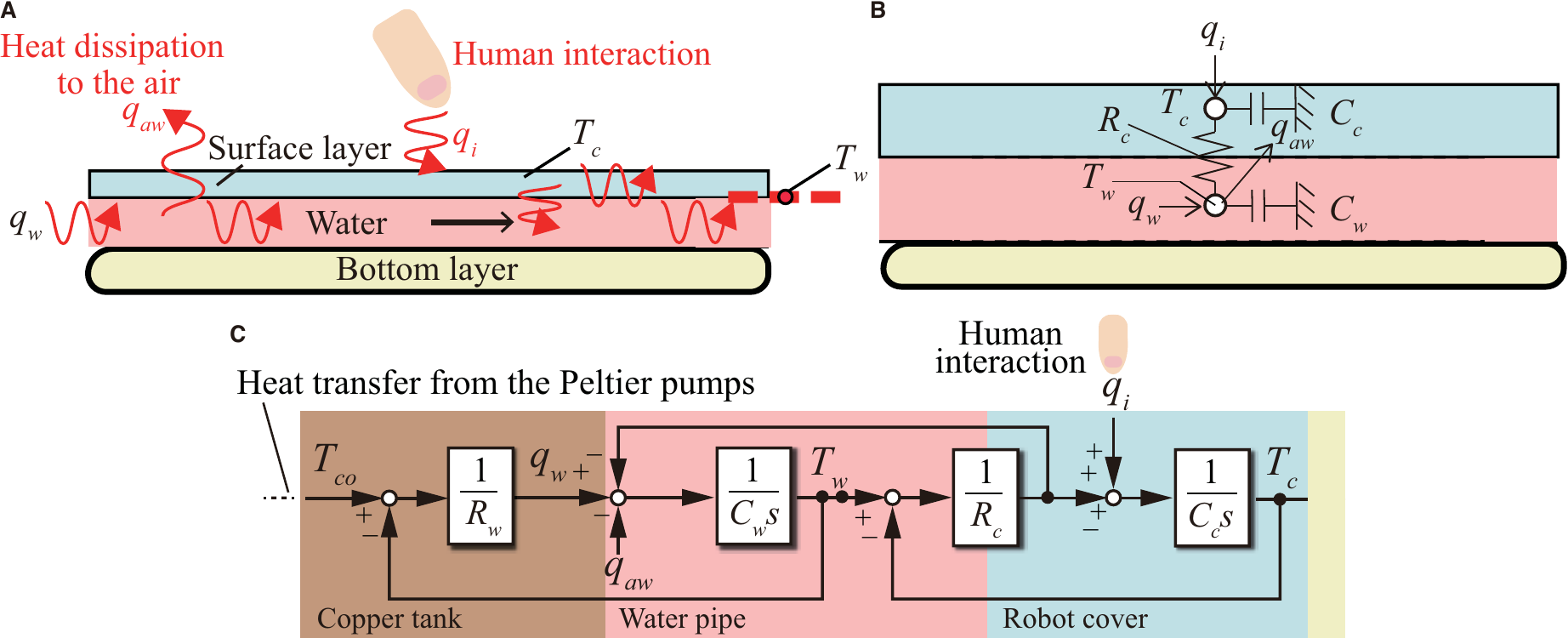}
\caption{Modeling of the circulating water system.}
\label{fig:modeling}
\end{center}
\end{figure}

\subsection{Heat flow observer for human interaction}
\label{sec:obs}
To detect altered thermal response of the robotic cover subsequent to human touch, heat flow is used. Indeed, heat flow behavior depends on the intrinsic (thermal) parameters of each material, and hence can be monitored to detect that a contact has been established. Moreover, the heat flow dynamic can (in theory) reveal if the contact is actually a human touch. Heat flow sensors exist, but the material by which they are composed biases heat conduction properties when it lies between two other materials (i.e. the robot cover and the human skin). In this paper, heat flow induced from a human touch on the cover is estimated from temperature rate information. The latter is obtained solely from thermocouples attached to the thermally conductive pipe (i.e. we do not use temperature measurement on the soft cover surface). 

To distinguish human touch, a heat flow observer is implemented. Our algorithm is based on the disturbance observer developed in~\cite{ohnishi1996tmech,morimitsu2015tie}, already used in many control applications. The estimation algorithm is simple; by monitoring the temperature of the copper box ($T_{co}$) and the water pipe ($T_w$) as observer inputs, heat flow from human contact $q_i$ is calculated based on the thermal model explained hereafter. 

\begin{figure}[t!]
\begin{center}
\includegraphics[width=0.8\columnwidth]{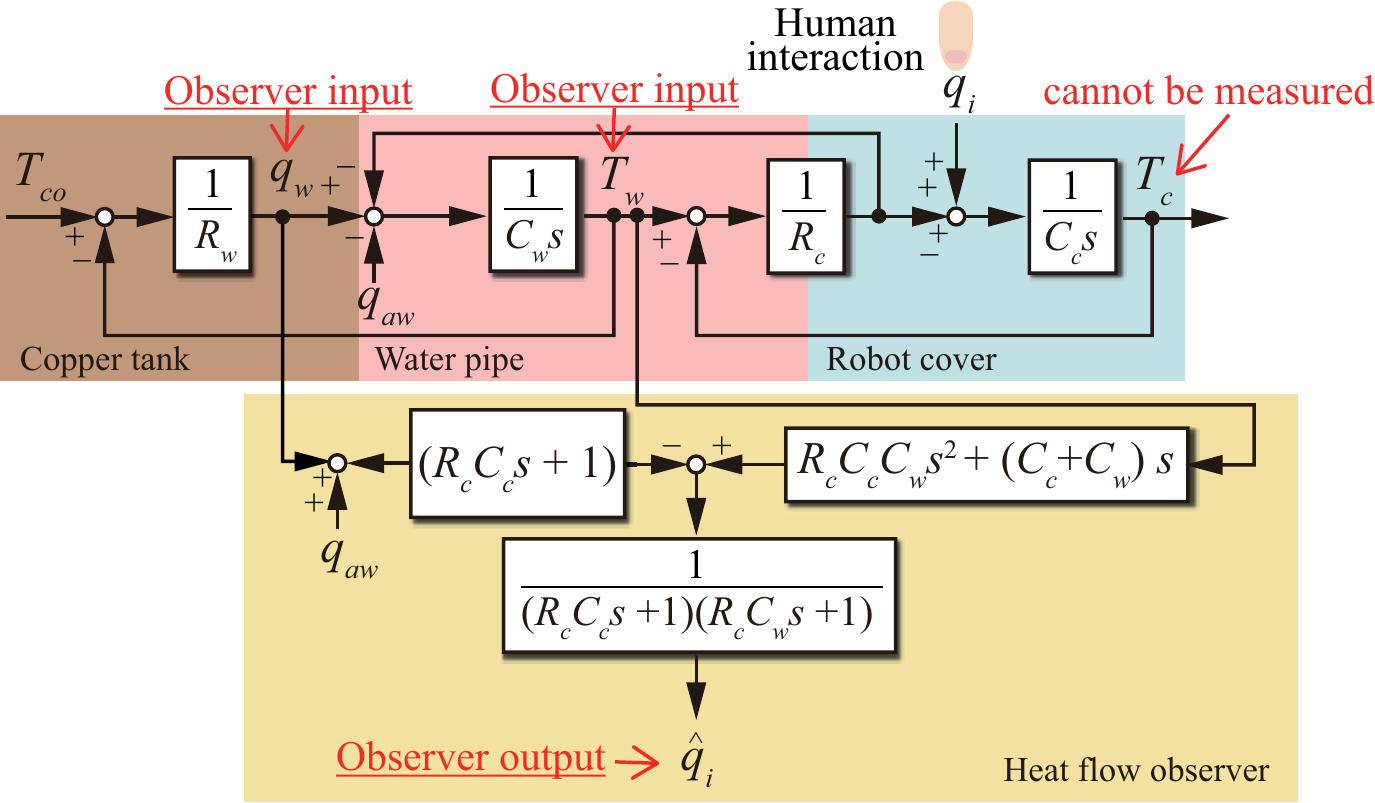}
\caption{The block diagram of the heat flow observer.}
\label{fig:block_observer}
\end{center}
\end{figure}

The thermal model of the circulating water system is shown in Fig.~\ref{fig:modeling}. It is derived from heat transfer modeling in~\cite{holman1990mcgraw}. 
Here, $T_w$, $T_c$, $R_w$, $R_c$, $C_w$, and $C_c$ stand for temperature, thermal resistance, thermal capacitance of the water pipe (index $_w$) and the soft cover (index $_c$). $T_{co}$ is the temperature of the copper tank, attached to the Peltier pumps. 
Heat flow $q_w$ convects through the circulating water, which is biased due to human-touch subsequent heat flow $q_i$ and heat loss $q_{aw}$. 
Based on the block diagram in Fig.~\ref{fig:modeling}C, the transfer function of the circulating water system is
\begin{align}
\frac{T_w}{q_w+q_{aw}} = \frac{R_cC_cs+1}{R_cC_wC_cs^2 + \left( C_w + C_c \right)s }, \label{eq:tr_qw}
\end{align}
where $s$ is the Laplace operator. Heat loss $q_{aw}$ is confirmed and estimated in preliminary experiments. We compensate for $q_{aw}$ to the transfer function input $q_w$ in~\eqref{eq:tr_qw}. 
The transfer function from $q_i$ to $T_w$ write as 
\begin{align}
\frac{T_w}{q_i} = \frac{ \left( R_cC_ws+1\right)\left( R_cC_cs+1\right) }{R_cC_wC_cs^2 + \left( C_w + C_c \right)s }. \label{eq:tr_qi}
\end{align} 
Using \eqref{eq:tr_qw} and \eqref{eq:tr_qi}, $T_w$ is expressed as
\begin{align}
T_w = \frac{\left(R_cC_cs + 1\right)}{R_cC_wC_cs^2 + \left( C_w + C_c \right)s } \left(q_w + q_{aw} \right) +
\frac{\left(R_cC_ws + 1\right)\left(R_cC_cs + 1\right)}{R_cC_wC_cs^2 + \left( C_w + C_c \right)s }q_i. \label{eq:observer_pre}
\end{align}
Based on eq.~(\ref{eq:observer_pre}), the observer for estimating $q_i$ is derived as
\begin{align}
\hat{q}_i = \frac{\left( R_cC_wC_cs^2 + \left( C_w + C_c \right)s \right)T_w- \left(R_cC_cs + 1\right) \left(q_w+q_{aw} \right) }{\left(R_cC_ws + 1\right)\left(R_cC_cs + 1\right)}. \label{eq:observer}
\end{align}
where $\hat{q}_i$ is the estimated value. 
By using eq.~(\ref{eq:observer}), a structure of the observer is expressed as shown in Fig.~\ref{fig:block_observer}. 
The observer output is directly affected by the parameter variation of~\eqref{eq:observer}. We conducted some identification tests by comparing to the model in Fig~\ref{fig:modeling}C with experimental results, deriving the exact parameters shown in Tab.~\ref{tab:identification}. 
As a result, the observer can extract only the heat flow from the contact object, regardless of the temperature that is displayed at the soft cover, as far as it is different from that of the object. 

	\begin{table}[t!]
		\begin{center}
		\caption{Identified parameters of the robotic cover system.} \label{tab:identification}
		\vspace{5pt}
		\begin{tabular}{c|c|c}\hline 
		Parameter  &Value &For\\
		\hline \hline 
	      $R_{\text{com}}C_{\text{com}}$ & 500 (heat), 450 (cool) seconds &Thermal display using $T_c$ (Experiment 1)\\  \hline
	      $L_d$ & 45 (heat), 30 (cool) seconds &Thermal display using $T_c$ (Experiment 1)\\  \hline
	       $R_a$ & 0.2 (heat), 0.3 (cool) K/W &Thermal display using $T_c$ (Experiment 1)\\  \hline
	       $R_{\text{com}}C_{\text{com}}$ & 500 (heat), 410 (cool) seconds &Thermal display using $T_w$ (Experiment 2) \\  \hline
	      $L_d$ & 45 (heat), 30 (cool) seconds &Thermal display using $T_w$ (Experiment 2)\\  \hline
	       $R_a$ & 0.3 (heat), 0.7 (cool) K/W &Thermal display using $T_w$ (Experiment 2)\\  \hline
	       $C_{co}$ &1152.57 J/K &Thermal exchange observer (Experiment 2)\\  \hline
	       $C_w$ &197.41 (heat), 182.79 (cool) J/K &Thermal exchange observer (Experiment 2) \\  \hline
	      $C_c$ &0.40 (heat), 0.10 J/K &Thermal exchange observer (Experiment 2)\\ \hline
               $R_{co}$ & 0.09 K/W &Thermal exchange observer (Experiment 2) \\  \hline
                $R_w$ & 6.00 (heat), 5.56 (cool) K/W &Thermal exchange observer (Experiment 2) \\  \hline
                $R_c$ & 120.12 (heat), 30.03 (cool) K/W &Thermal exchange observer (Experiment 2) \\ \hline
               $R_{aw}$ &2.1 K/W &Thermal exchange observer (Experiment 2) \\ \hline 
		\end{tabular}
		\end{center}
		\end{table}

\subsection{Control algorithm for thermal display}
\label{sec:control}
The closed-circuit water-circulation system has two controllable inputs: (i) the temperature of the Peltier pumps that are attached to the water tank; and (ii) the water flow rate, which can be changed by the voltage command of the water pump. A constant voltage applied to the water pump leads to a constant water flow rate and a linear model. In fact, given the size of the system and the slow dynamic of thermal heating and cooling, intermittent bang-bang type (max water flow, or no flow) command is sufficient. The heat transfer model of Fig.~\ref{fig:modeling}C can be approximated by summarizing multiple transfer layers (cooper water-tank, water-pipe...) as
	\begin{equation}
	T_c (s) = \frac{\exp(-L_ds)}{1+R_{\text{com}}C_{\text{com}}s} \left(T_p (s) - q_a \right),  \label{eq:conti}
	\end{equation}
where $s$, $T_p$, $T_c$, $R_{\text{com}}$, $C_{\text{com}}$, $q_a$, and $L_d$ stand for the Laplace operator, temperature of the Peltier pump and robotic cover, combined thermal resistance, combined thermal capacitance, heat loss at the surface of the robotic cover, and dead-time, respectively. 
A model preview controller (MPC) is used to regulate the temperature of the robotic cover. The cost function with additional constraints are chosen as
\begin{equation}
\min \sum_{i=1}^{H} \| \hat{T}_c(k+i|k) - T^{\text{cmd}}_{c}(k+i|k) \|^2_{W_1} + \sum_{i=0}^{H-1} \| \hat{T}_p(k+i|k)\|^2_{W_2} 
\label{eq:cost_function} 
\end{equation}
subject to
\begin{equation}
 T_c(k) = \frac{R_{\text{com}}C_{\text{com}}}{R_{\text{com}}C_{\text{com}}+t_s}T_c(k-1) + \frac{t_s}{R_{\text{com}}C_{\text{com}}+t_s}T_p(k-1-L_d/t_s)
\end{equation}
\begin{equation}
T_{\min}^{\text{th}} \leq T_p(k) \leq T_{\max}^{\text{th}},
\end{equation}
where $k$, $H$, $W_1$, $W_2$, $t_s$, $ \hat{T}_c$, $T_c^{\text{cmd}}$, $T_{\min}^{\text{th}}$, and $T_{\max}^{\text{th}}$ stand for the discrete time, prediction horizon, weight values of the cost function, sampling time, estimated value of $T_c$, temperature command of $T_c$, and minimum and maximum value of threshold of the calculated input, respectively. 
The first term of \eqref{eq:cost_function} works for tracking the temperature of the cover $T_c$ to its temperature command $T^{\text{cmd}}_{c}$, 
and the second one is for suppressing rapid change of the Peltier pump input. 
Here, the continuous model shown in~(\ref{eq:conti}) is converted to the discrete model. 
The same model is used for controlling $T_w$ with the observer explained in Sec.~\ref{sec:obs}. 

\section{Experimental results}
\label{sec:exp}
We achieved two kinds of experiments: (i) temperature control of the cover (Experiment 1: controlling $T_c$); and (ii) heat flow estimation without sensing the cover's surface temperature (Experiment 2: controlling $T_w$ and heat flow estimation). The identified parameters for both experiments are summarized in Table~\ref{tab:identification}. $R_{aw}$ is the thermal resistance between the water pipe temperature and ambient temperature, including $q_{aw}$ in Fig.~\ref{fig:block_observer}. These values were derived from experimental responses to step inputs and the model mentioned in Sections~\ref{sec:obs} and~\ref{sec:control}. 

\begin{figure}[t!]
\begin{center}
\includegraphics[width=\columnwidth]{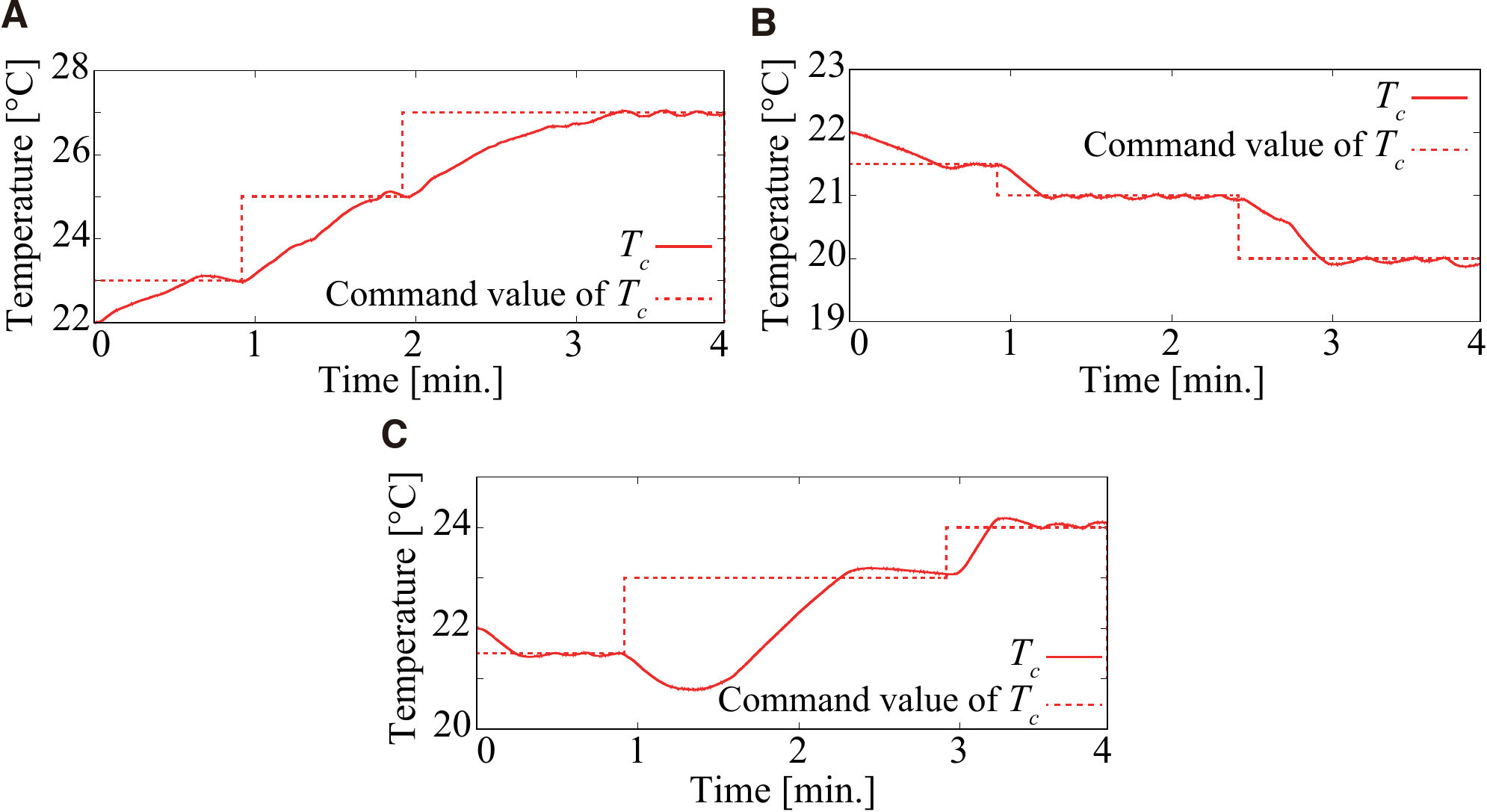}
\caption{Surface temperature control of the prototype cover (Experiment 1).}
\label{fig:temp_response}
\end{center}
\end{figure}

\subsection{Temperature control of the prototype cover (Experiment 1)}
In Experiment 1, the surface temperature of the prototype soft cover is controlled to be heated (with successive desired temperatures 23~$^\circ$C, 25~$^\circ$C, and 27~$^\circ$C) shown in Fig.~\ref{fig:temp_response}A, and then cooled (with successive desired temperature 21.5~$^\circ$C, 21~$^\circ$C, and 20~$^\circ$C) shown in Fig.~\ref{fig:temp_response}B. 
Figure~\ref{fig:temp_response}C shows the system responses heated to 23~$^\circ$C and 24~$^\circ$C after cooling to 21.5~$^\circ$C. The dotted lines plot desired temperatures, whereas plain lines plot the cover response when our control algorithm is applied. 

\begin{figure}[t!]
\begin{center}
\includegraphics[width=\columnwidth]{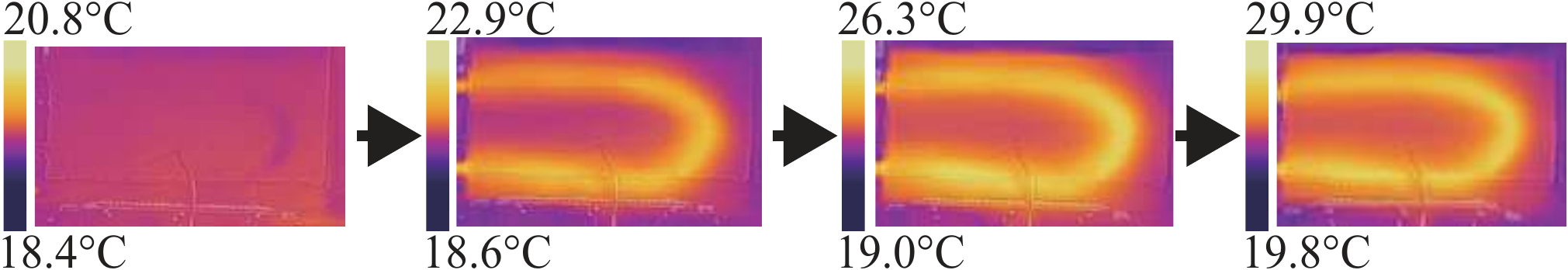}
\caption{Experimental results of Fig.~\ref{fig:temp_response}A monitored by an infrared camera (Experiment 1).}
\label{fig:thermography_heat}
\end{center}
\begin{center}
\includegraphics[width=\columnwidth]{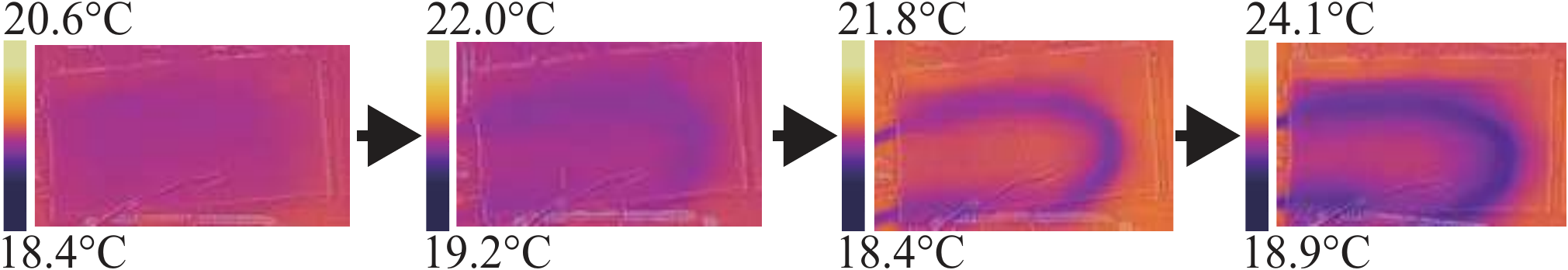}
\caption{Experimental results of Fig.~\ref{fig:temp_response}B monitored by an infrared camera (Experiment 1).}
\label{fig:thermography_cool}
\end{center}
\end{figure}

The Peltier pump's temperature command is generated so that the temperature of the robotic cover $T_c$ meets its desired $T_c^{\text{cmd}}$ temperature, and the response of the Peltier pump achieved subsequent to the controller. 
The circulating water is stopped whenever the response reaches its commanded desired value.
These experimental results show that the control system of the robotic cover is functioning correctly. 
When heating after cooling (or \emph{vice-versa}), it takes more time to meet the desired temperature because of the slow dynamics of thermal transfers (see Fig.~\ref{fig:temp_response}C). 
The main reason is related to the water's thermal capacitance, taking about 30 seconds to heat or cool for the whole circulating water. 
As additional evidence, Figures~\ref{fig:thermography_heat} and~\ref{fig:thermography_cool} show the surface of the prototype soft cover for the cases of Fig.~\ref{fig:temp_response}A and B monitored by an infrared camera. 

\begin{figure*}[t!]
        \begin{minipage}{0.5\hsize}
	\begin{center}
	\includegraphics[width=\hsize,clip]{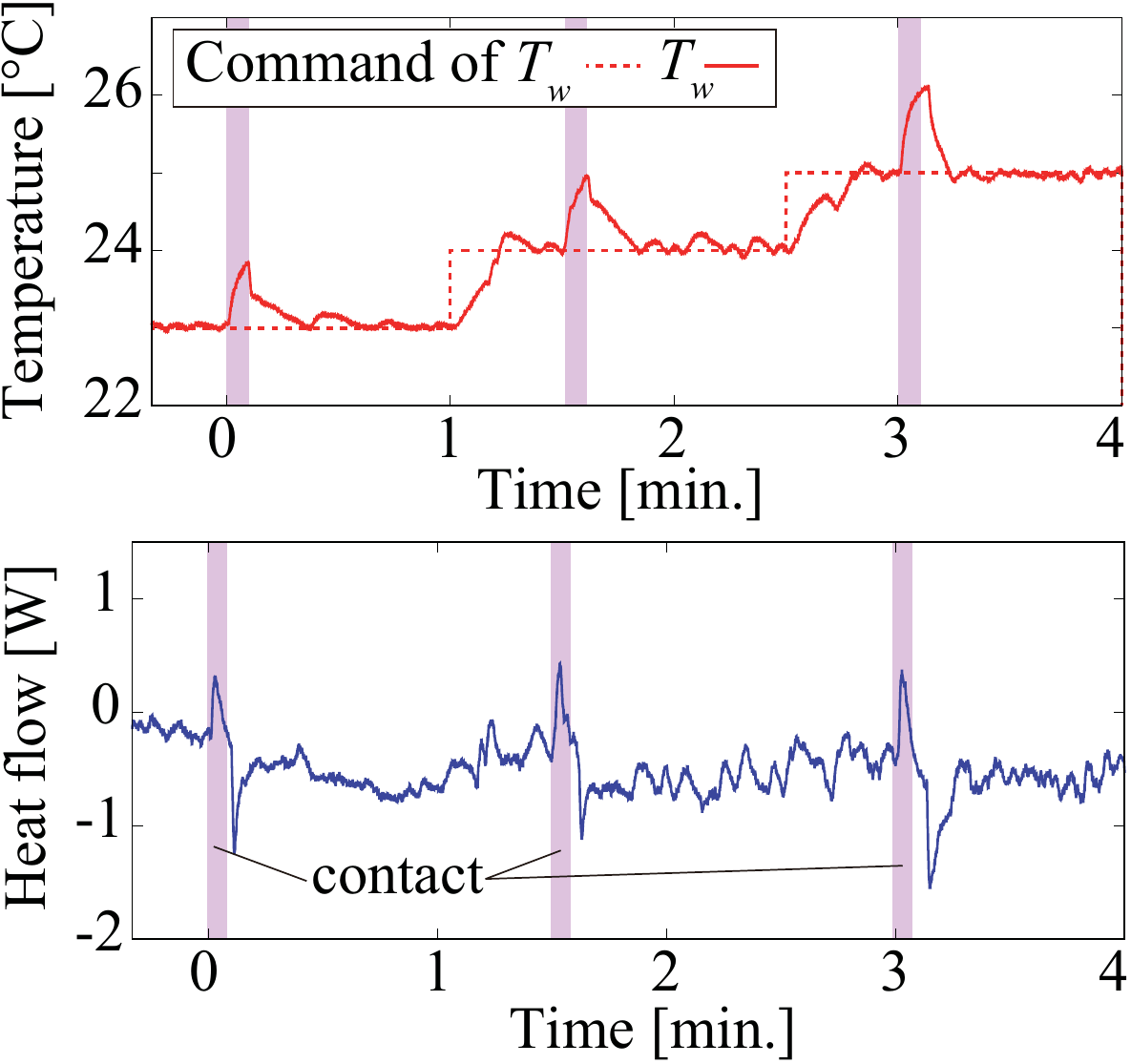}
	\end{center}
	\end{minipage}
	\begin{minipage}{0.5\hsize}
	\begin{center}
	\includegraphics[width=\hsize,clip]{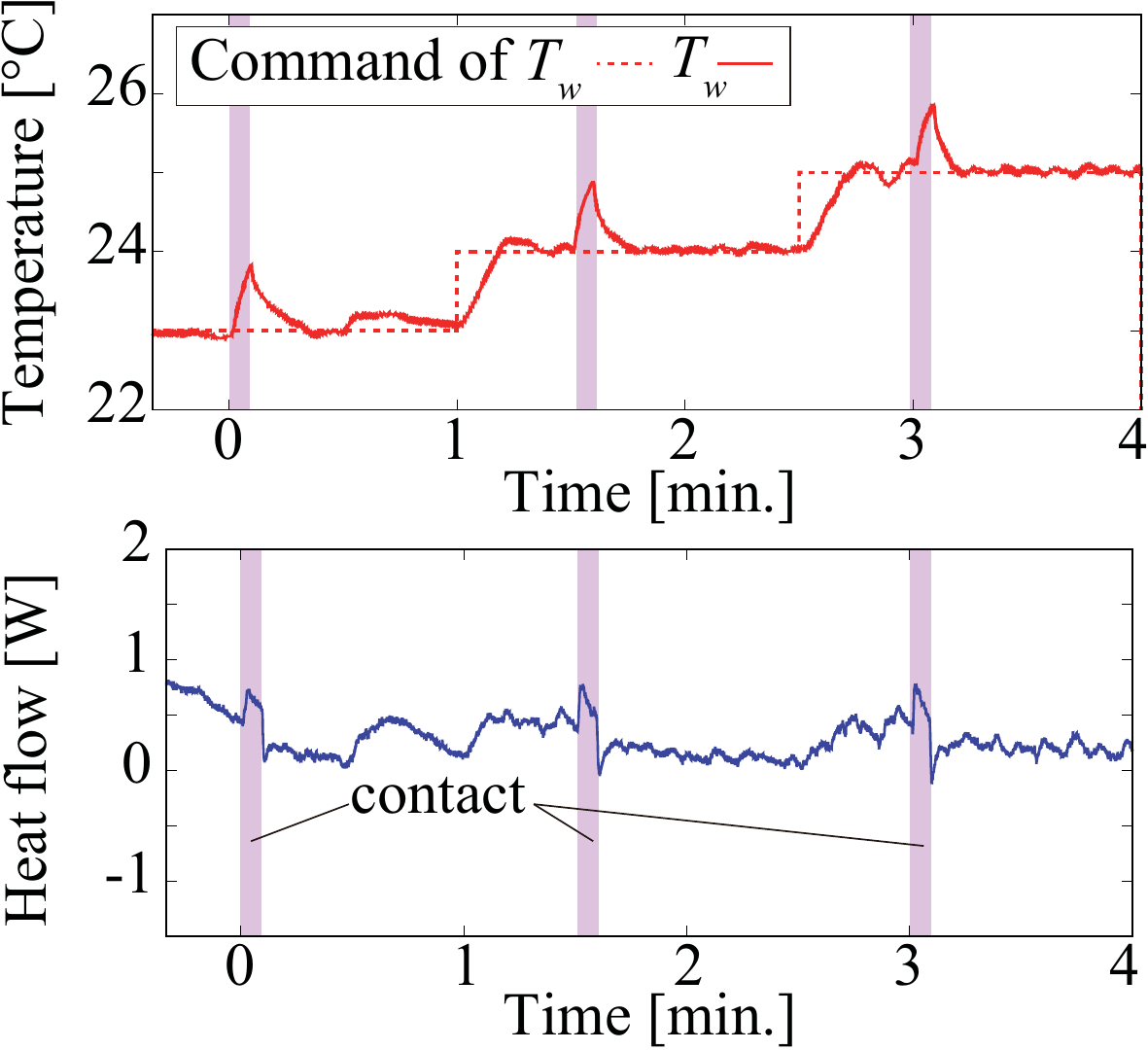}
	\end{center}
	\end{minipage}\\
	\begin{minipage}{0.5\hsize}
	\begin{center}
	\caption{Temperature and heat flow responses in grasping action (Experiment 2).}
	\label{fig:obs_grasp}
	\end{center}
	\end{minipage}
	\begin{minipage}{0.5\hsize}
	\begin{center}
	\caption{Temperature and heat flow responses in soft touch (Experiment 2).}
	\label{fig:obs_softtouch}
	\end{center}
	\end{minipage}\\
        \begin{minipage}{0.5\hsize}
	\begin{center}
	\includegraphics[width=\hsize,clip]{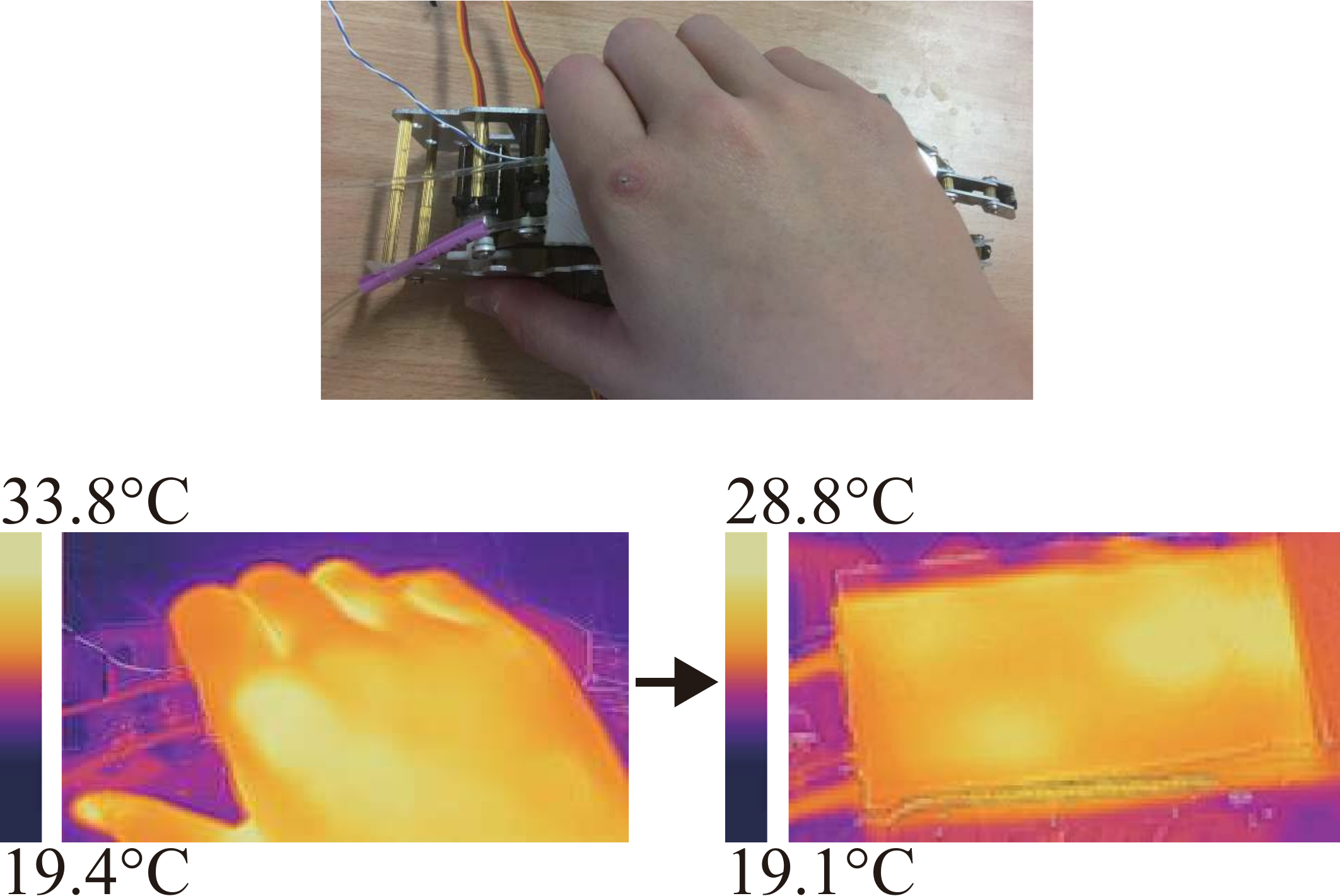}
	\end{center}
	\end{minipage}
	\begin{minipage}{0.5\hsize}
	\begin{center}
	\includegraphics[width=\hsize,clip]{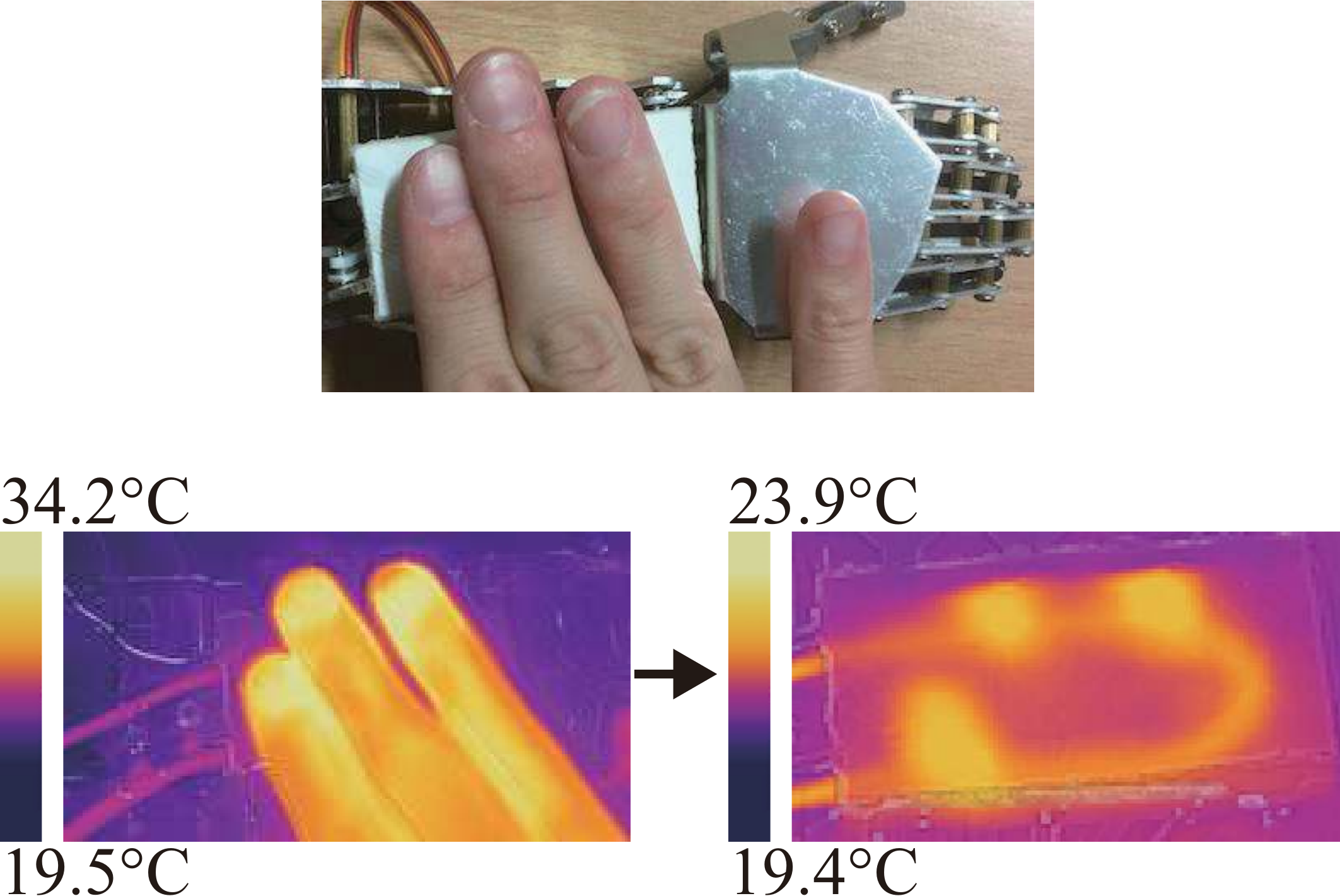}
	\end{center}
	\end{minipage}\\
	\begin{minipage}{0.5\hsize}
	\begin{center}
	\caption{Experimental results of Fig.~\ref{fig:obs_grasp} monitored by an infrared camera (Experiment 2).}
	\label{fig:obs_grasp_pic}
	\end{center}
	\end{minipage}
	\begin{minipage}{0.5\hsize}
	\begin{center}
	\caption{Experimental results of Fig.~\ref{fig:obs_softtouch} monitored by an infrared camera (Experiment 2).}
	\label{fig:obs_softtouch_pic}
	\end{center}
	\end{minipage}\\
\end{figure*}

\subsection{Heat flow estimation with temperature control (Experiment 2)}
In Experiment~2, we achieve heat flow estimation with temperature control. The estimated heat flow is made with the thermal exchange observer in Sec.~\ref{sec:obs}. We monitor two types of human contact with the soft cover: (i) grasp and (ii) soft touch, each lasting 5~sec. The conductive pipe temperature is controlled with successive steps 23, 24 and 25$^\circ$C for about 1~mn 30~sec each. Resulting temperature and estimated heat flow, together with infra-red camera monitoring of the soft cover, are shown in Figs.~\ref{fig:obs_grasp} to~\ref{fig:obs_softtouch_pic}, respectively. The heat flow observer is operating independently of the controlled water-pipe's temperature. Recall that the observer is based on temperature change of the water pipe due to a contact at the cover. The observer output depends on the amount of heat exchange when the human skin contacts the cover. For instance, the contact area for a grasp is wider relatively to a soft-touch (see Figs.~\ref{fig:obs_grasp} and~\ref{fig:obs_softtouch}). Yet, a limitation of the observer remains for extending it to contact localization and even to robust contact detection. Theoretically, the observed heat flow is zero when there is no contact, and increases or decreases only when contact occurs. However, the temperature response of the water pipe exhibit sometimes chattering around the desired temperature. This is because the water pump repeatedly switches between on and off when the response settles to its desired value. Subsequently, the observer output is not always zero when there is no contact. In practice, we can of course anticipate this chattering as we know when the pump switches on and off. However, the frequency of the switches is relatively high around the desired temperature and the contact can occur exactly at the moment of switch. The observer fails to detect touches or grasps lasting less than 5~sec as it has relatively low bandwidth. Indeed, the observer's output is too small to be distinguished from noise. We keep alive this problem and test other contact taxonomies such as holding and stroking with an integrated whole-scale system in future work.
\section{Conclusion and Future work}
\label{sec:conclusion}
We designed a new robotic cover that gathers both softness and thermal display (i.e. rendering) capabilities. On the bottom layer of the cover, we assembled thermally conductive pipes in which, water that can be heated or cooled, flows in a closed-circuit. We devised an integrated mechatronic system controller by means of a MPC that regulate the cover's surface temperature to any desired one. Moreover, we also noted that cover can also detect human contact through thermal features by implementing a thermal exchange observer. We hence obtain a dual soft thermal and sensing cover that can potentially be mounted on complex robots such as humanoids. 

\begin{acknowledgement}
This work is supported in part by JSPS Overseas Research Fellowships No.~201960463. 
\end{acknowledgement}

\bibliographystyle{spmpsci}
\bibliography{reference}

\end{document}